\documentclass[11pt]{article}

\usepackage{graphicx}

\usepackage{fullpage}
\usepackage[T3,T1]{fontenc}

\usepackage{amsmath,amssymb,amsfonts,amsbsy}
\usepackage{mathtools}
\usepackage{bm}

\usepackage{tikz}
\usetikzlibrary{calc,arrows.meta,decorations.pathreplacing}
\usepackage{tikz-3dplot}
\usepackage{caption}
\usepackage{subcaption}

\usepackage[numbers,square,sort&compress]{natbib}
\usepackage[normalem]{ulem}

\usepackage{hyperref}

\usepackage{xcolor}

\def\e{\mathrm{e}}

\def\d{\mathrm{d}}
\def\beq{\begin{equation}}
\def\eeq{\end{equation}}

\def\eps{\epsilon}

\def\P{{\sf P}}
\def\F{{\sf F}}

\def\C{{\sf C}}
\def\D{{\sf D}}
\def\Q{{\sf Q}}
\def\U{{\sf U}}

\def\AL{A^\textsc{l}}
\def\AR{A^\textsc{r}}

\newcommand{\eqn}[1]{(\ref{eqn:#1})}
\newcommand{\lab}[1]{\label{eqn:#1}}

\def\XXint#1#2#3{{%
  \setbox0=\hbox{$#1{#2#3}{\int}$}%
  \vcenter{\hbox{$#2#3$}}\kern-.5\wd0}}

\def\strutdepth{\dp\strutbox}
\def\nw#1{%
  \strut\vadjust{%
    \kern-\strutdepth
    \vtop to0pt{%
      \vss\hbox to\hsize{%
        \hskip\hsize\hskip5pt$\leftarrow$\hss\strut}}}%
  {\textcolor{teal}{\emph{#1}}}}

\title{Stationary distribution of active Brownian particles
in a channel domain}

\author{
Alexandra Tzella\textsuperscript{1},
Daniel Loghin\textsuperscript{1},
David J. Needham\textsuperscript{1}
and Jacques Vanneste\textsuperscript{2}
\\[1ex]
\small \textsuperscript{1}School of Mathematics,
University of Birmingham, Birmingham, UK
\\
\small \textsuperscript{2}School of Mathematics and Maxwell Institute,
University of Edinburgh, Edinburgh, UK
\\[1ex]
\small Corresponding author:
\href{mailto:a.tzella@bham.ac.uk}{a.tzella@bham.ac.uk}
}

\date{}

\begin{document}

\maketitle

\begin{abstract}
Self-propelled particles such as swimming micro-organisms often accumulate at boundaries. 
Active Brownian particles (ABPs), that is, particles moving at constant speed with an orientation undergoing rotational diffusion, provide a simple model. ABPs have a finite probability to  be on the boundary. Their probability density therefore includes both regular bulk and singular boundary components. Exchanges between these represent capture at the boundary and re-entry into the bulk.

We examine the distribution of ABPs in the stationary regime. We first obtain a similarity solution that describes the probability density near the turning points, where orientation is tangential to the boundary and particles re-enter the bulk. This shows that the bulk density is singular and the boundary density is non-differentiable. We then add spatial diffusion to the ABP model and investigate the  limit  $\eps \to 0$ of small spatial diffusivity. We use matched asymptotics to describe the regularising effect of a finite $\eps$, focusing on the vicinity of the turning points. We assess our predictions against numerical simulations using both lattice Markov chain  and finite-element discretisations.   
\end{abstract}

\medskip
\noindent\textbf{Keywords:}
active Brownian particles; boundary accumulation;
matched asymptotic expansions; singular perturbations;
bulk--boundary coupling; Fokker--Planck equations

\section{Introduction}
 Swimming microorganisms such as bacteria, algae and spermatozoa combine persistent 
self-propulsion with random changes in orientation to explore their surroundings. In homogeneous environments without directional cues, this active form of transport is well described at large scales by effective diffusion~\cite{Lovely_Dahlquist1975,Berg1993,CatesTailleur2013}. 
Recent experiments  link this large-scale behaviour to microscopic swimming statistics \cite{Kurzthaler_etal2024}.

Confinement changes this picture by introducing encounters with solid boundaries. Such encounters can lead to finite periods of residence on the boundary, which end when changes in orientation -- driven by rotational
diffusion, tumbling, or contact interactions -- allow microorganisms to return to the bulk \cite{LiTang2009,Drescher_etal2011,Kantsler_etal2013,Junot_etal2022}.
At population level, these finite residence times produce accumulation near the boundaries, where concentrations can greatly exceed those in the bulk~\cite{Rothschild1963,Berke_etal2008,Li_etal2011}.
Characterising the exchange between bulk and boundary is therefore central to
predicting the microorganism distribution.
  
While boundary encounters can involve complex hydrodynamic and steric effects~\cite{Lauga2020,Bechinger_etal2016}, boundary accumulation already arises in a minimal active Brownian particle (ABP) model combining constant-speed self-propulsion   with rotational diffusion~\cite{Romanczuk_etal2012}. 
In the absence of spatial diffusion,
ABPs encountering a boundary are simply assumed to remain on the boundary until rotational diffusion re-orients their velocity towards the bulk, enabling their re-entry into the bulk \cite{Li_etal2011,Lee2013}. 
ABPs therefore have a finite probability to be exactly   on the boundary. This results in a  probability density function (pdf) with two distinct components: a regular component supported on the bulk and a singular component supported on the boundary. The evolution of these two components is governed by coupled Fokker--Planck equations studied by Lee~\cite{Lee2013},  Wagner, Hagan and Baskaran~\cite{Wagner_etal2017}, and Fu, Perthame and Tang~\cite{FuPerthameTang2021}. 

The coupling term representing re-entry of boundary particles into the bulk takes the form of (Dirac) point sources at what we term turning points, that is, the points in the position--orientation space corresponding to boundary particles oriented tangentially to the boundary. The interpretation of this coupling term is delicate. As pointed out by Fu, Perthame and Tang~\cite{FuPerthameTang2021}, the 
structure of the solution at the turning points, specifically how singular it might be, is an open question. The question has both mathematical and physical implications. Mathematically, the answer determines how the equations governing ABP dynamics should be interpreted. Physically, it illuminates the nature of bulk--boundary exchanges of active particles.

In this paper, we examine the stationary distribution of ABPs   confined between two parallel walls, in the first instance without spatial diffusion (\S\ref{sec:ABP}). We derive a similarity solution near the turning points, expressed in terms of  Tricomi confluent hypergeometric function. This solution  reveals a singular bulk density and the square-root vanishing of the boundary density as they approach the turning points. This is in contrast with the linear vanishing that is implicitly assumed in the standard formulation of the confined ABP problem \cite{Lee2013,FuPerthameTang2021}. The singular behaviour shows that the stationary state results from the balance between two infinite fluxes from boundary to bulk and vice versa.

To understand this singular behaviour better, we then consider active Brownian particles in the presence of both rotational and spatial diffusion. We refer to this model as ABP$^+$ to distinguish it from  its zero spatial diffusivity counterpart, ABP.  
 ABP$^+$ under confinement has been studied in stationary and time-dependent settings~\cite{ElgetiGompper2013,Baouche_etal2026channel}. Impenetrability of the boundary is naturally imposed through a no-flux condition. Particles then have a zero probability to be exactly on the boundary and their pdf is smooth everywhere. For small spatial diffusivity $\eps \ll 1$, this can be thought of as resulting from the spreading out of the singular boundary probability of ABP over a thin boundary layer. 
This limiting picture emerges clearly from the asymptotic analysis of ABP$^+$ carried out by Berlyand et al.\ \cite{Berlyand_etal2020}. 

A complete asymptotic description of ABP$^+$ requires a separate treatment of the regions near the turning points. 
To provide this  description, we perform a detailed matched asymptotics analysis 
establishing consistent matching between the bulk, boundary-layer and turning-point regions (\S\ref{sec:ABP+}).
Our analysis clarifies the sense in which the confined ABP model can be regarded as a limit of the confined ABP$^+$ model as  $\eps \to 0$. This limit requires the swimming speed to be kept of order one. When this speed  is $O(\eps^{1/2})$, a different regime arises which we also examine (\S\ref{sec:weak}).

Throughout the paper, we assess our asymptotic predictions against numerical solutions obtained using a lattice Markov chain discretisation for the ABP problem and a finite-element discretisation for the ABP$^+$ problem. The Markov chain discretisation also provides a natural route to the very weak formulation of the   ABP problem considered by Fu, Perthame and Tang~\cite{FuPerthameTang2021} which is recovered in the continuum limit.

\section{ABP model}\label{sec:ABP}

\subsection{Formulation} \label{sec:ABPformulation}

The ABP model describes particles moving at constant speed $v$ in a direction that is changing randomly in time according to a Brownian process. In two dimensions, the position \((X,Y)\) and orientation \(\Phi\) of the particles is described by the SDEs
\beq
\d X=v\cos\Phi\,\d t,
\qquad
\d Y=v\sin\Phi\,\d t,
\qquad
\d\Phi=\sqrt{2}\,\d W,
\lab{ABP_SDE}
\eeq
where  \(W\) is a  Brownian (or Wiener) process on the circle. The  pdf  $p(x,y,\phi,t)$ 
for the position of particles in the three-dimensional position--orientation space is governed by the Fokker--Planck (or forward Kolmogorov) equation
\beq
p_t+v\cos\phi\,p_x+v\sin\phi\,p_y
= p_{\phi\phi}. 
\lab{ABP_FP}
\eeq

Equations\ \eqn{ABP_SDE} and \eqn{ABP_FP} are non-dimensionalised with the inverse of the rotational diffusivity \(D\) as reference time and
a length \(L\) as reference length.  Thus \(v\) is the swimming P\'eclet number,
\beq
v=\frac{\textrm{swimming speed}}{DL},
\eeq
measuring the strength of swimming advection relative to rotational diffusion.
 
\begin{figure}[t]
\centering
\includegraphics[width=0.5\textwidth]{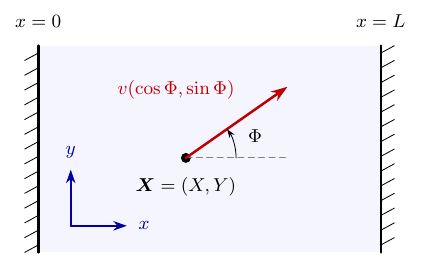}
\caption{ABP in a two-dimensional channel bounded by walls at $x=0$ and $x=L$. The particle is located at
$(X,Y)$ and swims with velocity
$v(\cos\Phi,\sin\Phi)$. For $\Phi=\pi/2$ and $\Phi=3\pi/2$, the particle swims parallel to the walls. }
\label{fig:particle_schematic}
\end{figure}

In the presence of spatial boundaries,  
\eqn{ABP_SDE} must be supplemented by a prescription
for the interaction of particles with the boundary. 
A common choice
suppresses the normal component of the particle velocity when it is directed out of the spatial domain, leaving it unchanged when it is directed into the domain. The tangential component is also left unchanged  \cite{Fu_etal2021}. For the infinite channel geometry represented in
Figure~\ref{fig:particle_schematic}, on which we focus, this prescription
reads
\beq
\d X = \begin{cases} 
v \left[\cos \Phi \right]_+ \, \d t & \textrm{for} \ \ X=0  \\
v \left[\cos \Phi \right]_- \, \d t & \textrm{for} \ \ X=1
\end{cases},
\quad
\d Y=v\sin\Phi\,\d t, \quad 
\d\Phi=\sqrt{2}\,\d W,
\lab{ABP_SDE_wall}
\eeq
where $[\cdot]_\pm$ denotes the positive and negative parts.
We take the channel width as the reference length \(L\), so $v$ is the ratio of  orientational persistence length to channel width.

\begin{figure}[t]
\centering
\includegraphics[width=0.75\textwidth]{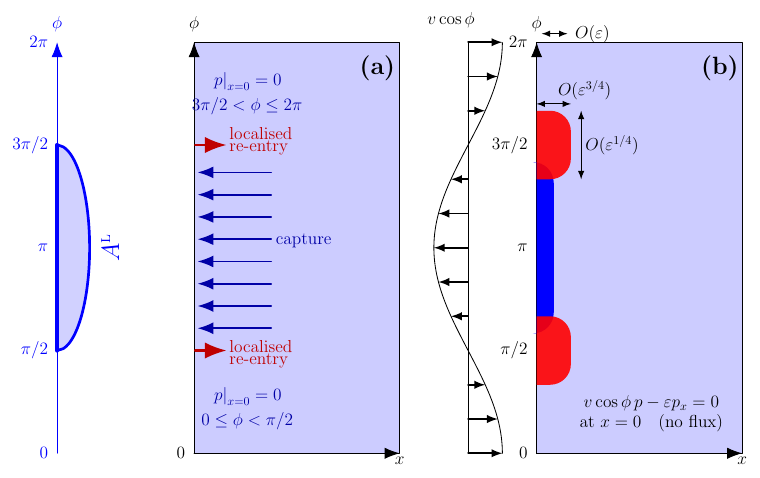}
\caption{ABP model and its regularisation by spatial diffusion
ABP\(^{+}\) near the left boundary \(x=0\).
(a) ABP:  the boundary density \(\AL(\phi)\) vanishes as a square root at
\(\phi=\pi/2\) and \(3\pi/2\); arrows indicate capture at the boundary and  re-entry into the bulk.
(b) ABP$^+$ with  \(\eps\ll1\) and \(v=O(1)\): different balances hold in the bulk,  \(O(\eps)\)
boundary layer (dark blue) and $O(\eps^{3/4}) \times O(\eps^{1/4})$ regions around the turning points $(0,\pi/2)$ and $(0,3\pi/2)$ (red).}
\label{fig:schematic}
\end{figure}

With \eqn{ABP_SDE_wall}, there is a finite probability for  particles to be on the boundaries at $x=0$ or $x=1$. The complete pdf is therefore written as 
\beq\lab{total_PDF}
\mathcal{P}(x,y,\phi,t)=p(x,y,\phi,t) + \AL(y,\phi,t) \delta(x) + \AR(y,\phi,t) \delta(x-1).
\eeq
Here \(p\) is the pdf for particles in the bulk of the domain, and $\AL$ and $\AR$ are the pdfs for particles on the left ($x=0$) and right ($x=1)$ boundaries. The bulk pdf $p$ is defined for \(x\in(0,1)\), \(y\in\mathbb{R}\) and $\phi \in [0, 2\pi)$. The boundary pdfs $\AL$ and $\AR$ are defined where particle advection is directed towards the boundary, that is,  $\AL$ is defined where \(\cos\phi<0\) and $\AR$ where \(\cos\phi>0\). They can be extended to $\phi \in [0,2\pi)$ by setting them to $0$ where undefined. All functions are then $2 \pi$-periodic in $\phi$. Panel (a) of figure \ref{fig:schematic} illustrates this set up near the left boundary. 

This paper is centred on the interactions between bulk and boundaries, that is, the coupling between $p$, $\AL$ and $\AR$. The $y$-dependence is incidental to this  and can be eliminated by considering the marginal probabilities obtained by integrating $p$, $\AL$ and $\AR$ over $y$. In the remainder we focus on these marginal probabilities.
This amounts to ignoring the $Y$ variable in \eqn{ABP_SDE} and \eqn{ABP_SDE_wall} and considering the position--orientation domain $(0,1) \times [0,2\pi)$. We abuse notation by writing the marginal pdfs simply as  $p$, $\AL$ and $\AR$.

Previous work~\cite{Lee2013,FuPerthameTang2021} formulates the evolution of \(p\), \(\AL\)
and \(\AR\) through the coupled system
\begin{subequations}
\lab{ABP_PDE}
\begin{align}
p_t + v \cos \phi \, p_x 
&= p_{\phi\phi} 
+\AL_\phi \, \delta(x)\left(\delta(\phi-\pi/2^+)
-
\delta(\phi-3\pi/2^-)
\right)
\nonumber  \\
& \qquad\quad + \AR_\phi \, \delta(x-1)\left(
\delta(\phi-3\pi/2^+)
-
\delta(\phi-\pi/2^-)
\right),
\lab{ABP_PDEa} \\
\AL_t 
&= \AL_{\phi \phi} - v \cos \phi \, p|_{x=0} \quad \textrm{for} \ \ 
\cos \phi <0, 
\lab{ABP_PDEb} \\
\AR_t 
&= \AR_{\phi \phi} + v \cos \phi \, p|_{x=1} \quad \textrm{for} \ \  
\cos \phi >0. 
\lab{ABP_PDEc}
\end{align}
The system   is supplemented by the  boundary conditions  
\begin{align}
p|_{x=0} &= 0 \quad \textrm{for} \ \ \cos\phi>0,
\qquad
\AL = 0 \quad \textrm{for} \ \  \cos\phi\geq0,
\lab{ABP_PDEd}\\
p|_{x=1} &= 0 \quad \textrm{for} \ \ \cos\phi<0,
\qquad
\AR = 0 \quad \textrm{for} \ \ \cos\phi\leq0.
\lab{ABP_PDEe}
\end{align}
\end{subequations}
Total probability, including the boundary-supported
contributions, is conserved and  normalized to $1$. 
 
The interpretation of   system \eqn{ABP_PDE} is seemingly straightforward. In the bulk, the particle density $p$ evolves according to the Fokker--Planck equation \eqn{ABP_FP}, recovered from \eqn{ABP_PDEa}. 
Particles are captured on the boundaries when their velocity is directed  towards the boundaries. The probability flux $-v\cos\phi\,\left.p\right|_{x=0}$ acts as a source for the boundary density $\AL$, and the probability flux  $v\cos\phi\left.p\right|_{x=1}$ acts as a source for $\AR$, as described by \eqn{ABP_PDEb}--\eqn{ABP_PDEc}.

The terms proportional to $\delta(x)$ and $\delta(x-1)$ in \eqn{ABP_PDEa} describe the re-entry of boundary particles into the bulk. This re-entry is localised at the four turning points $(0,\pi/2)$, $(0,3\pi/2)$, $(1,\pi/2)$ and $(1,3\pi/2)$, corresponding to particles whose velocity is tangent to the boundary.
However, interpreting the re-entry terms as localised sources requires their strengths, given by the (one-sided) derivatives \(\AL_\phi\) and \(\AR_\phi\) at  $\phi=\pi/2$ and  $3\pi/2$, to be finite. We show below that this is not the case for the stationary solution.

From now on, we restrict our discussion to the stationary solution, 
setting $p_t=\AL_t=\AR_t=0$.
The stationary equations are homogeneous with a normalisation constraint  (given explicitly in \eqn{ABP_PDEf} below). 
The stationary pdfs  satisfy the symmetries $p(x,\phi)=p(1-x,\phi+\pi)=p(x,-\phi)$ and $\AR(\phi)=\AL(\phi+\pi)$ which imply $p_\phi(x,0)=p_\phi(x,\pi)$=0. 
We do not exploit these symmetries.

\subsection{Turning points and re-entry}\label{subsec:reentry}

A robust way of understanding the ABP model with boundary is as the limit of a continuous-time Markov chain on a lattice as the lattice spacing tends to $0$. The lattice Markov chain, which we describe in Appendix \ref{app:MC}, also provides an effective discretisation which we use to compute the stationary $p$, $\AL$ and $\AR$. 

\begin{figure}[t]
\centering
\includegraphics[width=0.49\linewidth]
    {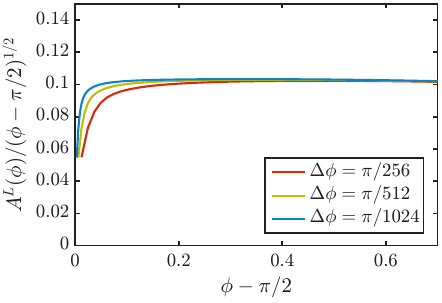}%
\hspace{0.01\linewidth}%
\includegraphics[width=0.49\linewidth]
    {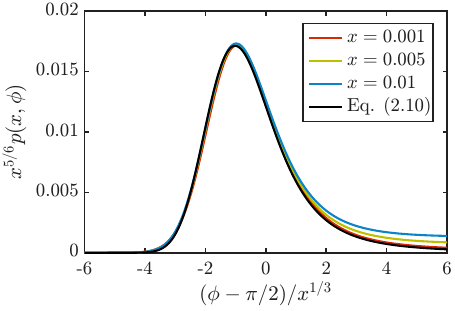}
\caption{Stationary solution of  ABP with $v=1$ obtained using the
lattice Markov chain in \eqn{MC}
compared with the asymptotic predictions
\eqn{assumption} and \eqn{self-similar}.
Left:  rescaled boundary density $\AL(\phi)/(\phi-\pi/2)^{1/2}$ for three angular lattice spacings
$\Delta\phi$. 
Right: rescaled bulk density $x^{5/6}p(x,\phi)$
plotted against the similarity variable
$(\phi-\pi/2)/x^{1/3}$ at three values of $x$
using $\Delta\phi=\pi/1024$, and compared with a
best-fit multiple of the similarity solution $\F$ in \eqn{F}.
The comparison is made in the turning-point region
$\phi-\pi/2=O(x^{1/3})$ with $ x\ll1$.
}
\label{fig:asymptotics}
\end{figure}

Figure \ref{fig:asymptotics}(a) shows the discrete approximation $\AL$ obtained in this way for three different $\Delta \phi$, the lattice spacing in the $\phi$-direction. The figure focuses on the vicinity of the turning point $\phi=\pi/2$ and displays $\AL$ rescaled by $(\phi-\pi/2)^{1/2}$. The observed convergence to a constant away from an $O(\Delta \phi)$ neighbourhood of the turning point strongly suggests that, in the continuum limit, $\AL \propto (\phi-\pi/2)^{1/2}$ and therefore $\AL_\phi \propto(\phi-\pi/2)^{-1/2}$ diverges as $\phi \to \pi/2$. A similar behaviour applies to the other three turning points.

We explain this behaviour  by considering the solution to \eqn{ABP_PDE} near the turning point \(x=0\), \(\phi=\pi/2\). The corresponding behaviour at the other three turning points follows. 
Using \(\cos\phi\sim-(\phi-\pi/2)\) as
\(\phi\to\pi/2\),  \eqn{ABP_PDE} reduces locally to
\begin{subequations}\lab{ABP0_local_bulkboundary}
\begin{align}
-v(\phi-\pi/2)\,p_x
&=
p_{\phi\phi}
\quad \textrm{for} \ \ 
x>0,
\qquad
p|_{x=0} = 0  \quad 
\textrm{for}
 \  \phi<\pi/2,
\lab{ABP0_local_bulk}
\\
\AL_{\phi\phi}
&=
-v(\phi-\pi/2)\,p|_{x=0}
 \quad 
\textrm{for} \ \ 
\phi>\pi/2,
\qquad
\AL =0 
 \quad 
\textrm{for} \ \
\phi \le \pi/2.
\lab{ABP0_local_wall}
\end{align}
\end{subequations}

The bulk equation \eqn{ABP0_local_bulk} is, up to rescaling, the
stationary Kolmogorov equation~\cite{Kolmogorov1934}, supplemented here
by a  homogeneous boundary condition for $\phi < \pi/2$. 
Both the equation and  boundary condition are invariant under the scaling
$(x,\phi-\pi/2)\mapsto(\lambda^3x,\lambda(\phi-\pi/2))$.
This suggests seeking a self-similar solution  
of the form
\beq\lab{self-similar}
p(x,\phi)
= 
\frac{\C}{4v}
\left(\frac{v}{9}\right)^{5/6}\,
x^b\,\F(\eta)
\quad \textrm{with} \ \ 
\eta=\frac{\phi-\pi/2}{x^{1/3}},
\eeq
where $\C$ is a constant and the exponent \(b\) is to be determined.
The prefactor involving $v$ is introduced for later convenience. 
Substituting into \eqn{ABP0_local_bulk} gives
 \begin{equation}\lab{Kummer}
 \F_{\eta \eta}-\frac{v}{3}\eta^2\F_\eta+b v\eta \F=0.
 \end{equation}
 The change of variable \(z=v\eta^3/9\) reduces this equation to
Kummer's confluent hypergeometric equation
\cite[Eq.~(13.2.2)]{DLMF}. The  two independent large-|\(\eta\)|
behaviours the solutions of \eqn{Kummer} are algebraic and exponentially decaying.

For \(\eta<0\), the algebraic behaviour is incompatible with the zero boundary
condition at \(x=0\), so we retain the exponentially decaying solution.
We therefore set
\(
z=-v\eta^3/9>0
\)
and write
\(
\F=e^{-z}\widetilde{\F}.
\)
Substitution gives Kummer's confluent hypergeometric equation for
\(\widetilde{\F}\), whose solution may be expressed in terms of the 
Tricomi's confluent hypergeometric function (or 
confluent hypergeometric function of the second kind) \(\U\), giving
 \begin{subequations}\lab{F}
\begin{equation}\lab{F-}
\F(\eta)
=
\C^- \e^{v\eta^3/9}
\U\left(b+\frac23,\frac23,-\frac{v\eta^3}{9}\right),
\end{equation}
where $\C^-$ is a constant. 
For \(\eta>0\), we exclude the exponentially growing solution and
retain the Tricomi solution, so
\begin{equation}\lab{F+}
\F(\eta)
=
 \U\left(-b,\frac23,\frac{v\eta^3}{9}\right),
\end{equation}
\end{subequations}
where we set the prefactor to $1$ using the arbitrariness of $\C$ in \eqn{self-similar}.

Eqs.\ \eqn{F} give the solution to \eqn{Kummer} on the two half lines in term of Tricomi
functions, both with positive argument.
Since \(\eta=0\) is an ordinary point of \eqn{Kummer}, the two
half-line solutions   must be restrictions
of a single analytic solution. Hence \(\F\) and \(\F_{\eta}\)
are continuous at \(\eta=0\).
Using \cite[Eq.~(13.2.42)]{DLMF} this yields
\begin{equation}\lab{continuity_conditions}
\C^-
\frac{\Gamma(1/3-b)}{\Gamma(b+1)} =  1
\quad \textrm{and} \quad 
\C^- 
\frac{\Gamma(-b)}{\Gamma(2/3+b)} = -  1.
\end{equation}
Eliminating $\C^-$ gives
\(\Gamma(\tfrac13-b)\Gamma(\tfrac23+b)=-\Gamma(-b)\Gamma(1+b)\); by
\(\Gamma(\tfrac13-b)\Gamma(\tfrac23+b)=\pi/\sin\pi(\tfrac13-b)\) and
\(\Gamma(-b)\Gamma(1+b)=-\pi/\sin\pi b\), this reduces to
\(\sin\pi(\tfrac13-b)=\sin\pi b\), hence \(b=\tfrac16+n\),
\(n\in\mathbb Z\).
The boundary condition in   \eqn{ABP0_local_bulk} requires that \(b<0\), so \(n\leq-1\). The requirement that
\(p\geq0\) imposes that the solution \(\F\)  be sign-definite.
For positive argument, \(\U(-b,\tfrac23,x)\) has no zeros when
\(b<0\) \cite[Eq.~(13.9.15)]{DLMF}, while
\(\U(b+\tfrac23,\tfrac23,x)\) has no zeros only when
\(b\geq-1\) \cite[Eq.~(13.9.14)]{DLMF}.
Thus \(-1\leq b<0\) which selects $n=-1$ in the sequence \(b=1/6+n\) so that
\begin{equation}\lab{similarity-aid}
b=-5/6
\quad \textrm{and} \quad
\C^-=6,
\end{equation}
where the last relation uses \(\Gamma(7/6)= \Gamma(1/6)/6\). 

The  similarity solution \eqn{self-similar}--\eqn{F} with
\eqn{similarity-aid} is a central result of this paper.  It captures the exact form of the stationary pdf near the turning points uniformly in $x$ and $\phi-\pi/2$, with $\eta=O(1)$ the relevant distinguished regime.  
We compare this similarity solution with a numerical solution of the stationary solution in figure~\ref{fig:asymptotics}(b). The numerical solution is computed using the lattice Markov chain model detailed in Appendix \ref{app:MC}.
The solution uses a uniform grid in $\phi$ and, motivated by the similarity variable $\eta=(\phi-\pi/2)/x^{1/3}$, a  non-uniform grid in $x$ with $\Delta x \propto x^{1/3}$ near the boundary. Robust results are thus 
 obtained without an exceedingly large number of lattice points.
The similarity solution  is recovered
for $x$ small (but large enough that discretisation effects are negligible,
say from the $10$th lattice point on). For $x>0.001$, the discrete solution departs from the
similarity profile, reflecting the local nature of the approximation.

To characterise the singular behaviour of the solution near the turning point,
we use $\U(a,b,z)\sim z^{-a}$ as $z\to\infty$
\cite[Eq.~(13.7.3)]{DLMF} to obtain
\begin{equation}\lab{F_lim}
\F(\eta)
\sim 
\left(\frac{v}{9}\right)^{1/6}
\begin{cases}
6
(-\eta)^{1/2}
\e^{{v\eta^3}/{9}}
\quad
&\textrm{as} \ \ \eta\to-\infty.\\
\eta^{-5/2}
\quad
&\textrm{as} \ \ \eta\to+\infty.
\end{cases}
\end{equation}
 For fixed $\phi<\pi/2$, letting $x\to 0^+$ gives $\eta\to-\infty$, and the exponentially decaying branch in \eqn{F_lim} is consistent with the zero boundary condition at $x=0$.  For fixed $\phi>\pi/2$, letting $x\to 0^+$ gives $\eta\to+\infty$. Using \eqn{F_lim}, the powers of $x$ cancel, so that $\lim_{x\to 0^+}p(x,\phi)=p(0,\phi)$ is finite for fixed $\phi>\pi/2$, although it becomes singular as  $\phi\to \pi/2^+$.
Specifically,
\begin{equation}\lab{consequence}
p(0,\phi)\sim \frac{\C}{4 v} \left(\phi-\frac{\pi}{2}\right)^{-5/2} \quad \text{as} \ \  \phi\to\frac{\pi}{2}^{+}.
\end{equation}
Substituting \eqn{consequence} into \eqn{ABP0_local_wall},
integrating twice and using \(A(\pi/2)=0\) gives the leading
square-root behaviour
\begin{equation}\lab{assumption}
\AL(\phi)
\sim
\C
\left(\phi-\frac{\pi}{2}\right)^{1/2}
\quad
\text{as} \ \  \phi\to\frac{\pi}{2}^{+},
\end{equation}
in agreement with 
Figure~\ref{fig:asymptotics}(a). 

\subsection{Interpretations of  the ABP model}
Since the derivatives  $\AL_\phi$ and $\AR_\phi$ of the boundary pdfs diverge to infinity at the turning points, 
the finite-strength interpretation of the   re-entry sources in \eqn{ABP_PDEa} breaks down. The system \eqn{ABP_PDEa} therefore needs a re-interpretation which we now discuss.

Collecting the leading-order bulk, boundary-layer and turning-point
descriptions leads to a refined formulation of the ABP stationary problem, in terms
of the bulk density \(p(x,\phi)\) and the left- and right-wall
densities \(\AL(\phi)\) and \(\AR(\phi)\).  The governing equations are
\begin{subequations}\lab{ABP0_bulkboundary_model}
\begin{align}
v\cos\phi\,p_x
&=
p_{\phi\phi},
\qquad
0<x<1,
\quad
0\leq\phi<2\pi,
\lab{ABP0_bulk_interior}
\\
\AL_{\phi\phi}
&=
v\cos\phi\,p|_{x=0},
\qquad
\cos \phi < 0, 
\lab{ABP0_leftwall}
\\
\AR_{\phi\phi}
&=
- v\cos\phi\,p|_{x=1},
\qquad
\cos \phi > 0.
\lab{ABP0_rightwall}
\end{align}
These are supplemented by the boundary conditions
\begin{align}
p|_{x=0}
=
0
\quad  \textrm{for} \ \ 
\cos\phi>0,
\qquad
p|_{x=1}=0
\quad \textrm{for} \ \ 
\cos\phi<0,
\lab{ABP0_inflow_bulkboundary}
\\
\AL(\pi/2)=\AL(3\pi/2)=\AR(\pi/2)=\AR(\pi/2)=0.
\lab{ABP0Abc}
\end{align}
The system is completed by specifying the singular behaviour   
\beq
p(x,\phi)
\sim 
 \frac{\C}{4v}
\left(\frac{v}{9}\right)^{5/6}
\F\!\left(
\frac{\phi-\pi/2}{x^{1/3}}
\right)
\quad \textrm{as} \ \
(x,\phi)\to
\left(0,\frac{\pi}{2}\right),
\lab{ABP0_turning_bulkboundary}
\eeq
with $\F$ defined in \eqn{F}. 
The behaviour at the other turning
 points \((0,3\pi/2)\), \((1,\pi/2)\) and \((1,3\pi/2)\) is similar.
 \end{subequations}

The system \eqn{ABP0_bulkboundary_model} can be solved by first solving \eqn{ABP0_bulk_interior} with boundary conditions \eqn{ABP0_inflow_bulkboundary} and imposing the singular behaviour \eqn{ABP0_turning_bulkboundary} for an arbitrary $\C$. Eqs.\ 
\eqn{ABP0_leftwall} and \eqn{ABP0_rightwall} can then be solved with the boundary conditions \eqn{ABP0Abc}. The constant $\C$ is finally determined from the normalisation condition 
\begin{equation} 
\int_0^1\int_0^{2\pi}
p\,\d x  \d\phi
+
\int_{0}^{2\pi}
\AL\,\d\phi  
+
\int_0^{2\pi}
\AR\,\d\phi 
=
1,
\lab{ABP_PDEf}
\end{equation}
The first integral is finite despite the turning-point singularity of the bulk density.
This is because 
\(p\,\mathrm dx \, \mathrm d\phi\propto x^{-1/2}\F(\eta)\,\mathrm dx\,\mathrm d\eta\),
and both $x^{-1/2}$ and $F(\eta)$ are integrable (see  \eqn{F_lim}).

An alternative interpretation of the stationary ABP model is in terms of a very weak formulation.
The easiest way to obtain this is as the limit of the corresponding formulation of the lattice Markov chain model \eqn{MC}. 
As detailed in appendix \ref{app:MC}, this leads to
\beq
\int_0^1  \int_0^{2\pi}
p(x,\phi)
\left(
v\cos\phi\,q_x+q_{\phi\phi}
\right)
\, \d\phi\,\d x
+
\int_0^{2\pi}
\AL(\phi)q_{\phi\phi}|_{x=0} \,\d\phi
+
\int_0^{2\pi}
\AR(\phi)q_{\phi\phi}|_{x=1} \,\d\phi = 0,
\lab{eq:FV_continuum_weak}
\eeq
where $q$ is an arbitrary, sufficiently smooth \(2\pi\)-periodic test function on \([0,1]\times[0,2\pi)\). This very weak formulation \eqn{eq:FV_continuum_weak} is derived directly
from the PDE system \eqn{ABP_PDE} by Fu, Perthame and Tang~\cite{FuPerthameTang2021}.
It can be expressed in terms of the total pdf  $\mathcal{P}$ as
\begin{subequations}
\lab{eq:FV_continuum_weak2}
\beq
\int_0^1\int_0^{2\pi}
\mathcal{P}(x,\phi)
\left(v\cos\phi\,q_x+q_{\phi\phi}\right)
\,\d\phi\,\d x=0, \lab{L}
\eeq
with associated boundary conditions
\beq
\left.q_x\right|_{x=0}=0 \quad \text{for} \ \ \cos\phi<0,
\qquad
\left.q_x\right|_{x=1}=0\quad \text{for} \ \ \cos\phi>0. \lab{Lbc}
\eeq
\end{subequations}
The generator $\mathcal{L}$ of the ABP process can be identified from  \eqn{eq:FV_continuum_weak2} and the definition  $\langle \mathcal{P}_t,q \rangle = \langle \mathcal{P}, \mathcal{L} q \rangle$ with $\langle \cdot, \cdot \rangle$ the $L_2$ inner product (e.g.\ \cite{BouRabeeHolmesCerfon2020}),  as
 $\mathcal{L} =  v \cos \phi \, \partial_{x} + \partial_{\phi \phi}$ with boundary conditions \eqn{Lbc}.
Note that $q=1$ cancels the left-hand side of \eqn{L} for arbitrary $\mathcal{P}$, reflecting probability conservation.

\section{ABP$^+$} \label{sec:ABP+}
 \subsection{Formulation}

The results of \S\ref{sec:ABP} show that the ABP model is singular: in steady state, the bulk pdf $p$ and the derivatives $\AL_\phi$ and $\AR_\phi$ of the boundary pdfs diverge to infinity at the turning points. We expect this to be regularised by physical effects neglected for ABPs. In this section we examine one such effects namely a weak spatial diffusion. We show how the  small spatial-diffusivity limit recovers the ABP results while producing a smoothed-out re-entry region around the turning points.

We refer to the model of active Brownian particles including spatial diffusion as  ABP\(^{+}\) model. The dimensionless governing SDEs are 
\beq
\d X
=
v\cos\Phi\,\d t+\sqrt{2\eps}\,\d W',
\qquad
\d Y
=
v\sin\Phi\,\d t+\sqrt{2\eps}\,\d W'',
\qquad
\d\Phi
=
\sqrt{2}\,\d W,
\lab{ABPplus_SDE}
\eeq
where \(W'\), \(W''\) and \(W\) are independent Wiener processes.
The dimensionless spatial diffusivity
\begin{equation}
\eps=\frac{\text{spatial diffusivity}}{DL^2}
\end{equation}
measures the relative importance of spatial and rotational diffusion.
We focus on the limit $\eps \to 0$.

The  Fokker--Planck equation corresponding to \eqn{ABPplus_SDE} is 
\begin{subequations}\lab{ABPdiff}
\beq
p_t+v\cos\phi\,p_x+v\sin\phi\,p_y
=
p_{\phi\phi}+\eps(p_{xx}+p_{yy})
\lab{ABPdiffFB}
\eeq
posed on $(0,1) \times [0, 2\pi)$.  
The boundary conditions are the straightforward no-flux conditions 
\beq
v\cos\phi\,p-\eps p_x=0
\quad \textrm{at} \ \
x=0,1, \quad 
 0\leq\phi<2\pi,  
\lab{ABPdiffBC}
\eeq
\end{subequations}
together with \(2\pi\)-periodicity in \(\phi\). The ABP$^+$ problem \eqn{ABPdiff} shares the  symmetries of the ABP problem mentioned in \S\ref{sec:ABPformulation}.

As in \S\ref{sec:ABP}, we focus on the stationary solution marginalised over $y$ and therefore set $p_t=p_y=0$.
The  ABP\(^{+}\) problem  then reduces to 
\begin{subequations}\lab{ABP1D_tot}
\begin{equation}
v\cos\phi\,p_x
=
p_{\phi\phi}+\eps p_{xx}
\quad \textrm{with} \ \
v\cos\phi\,p
=
\eps p_x
\ \  \textrm{at} \ \ x=0,1,
\lab{ABP1D}
\end{equation}
with  normalisation
\beq
\int_0^{2\pi}\int_0^1 p(x,\phi)\,\d x\,\d\phi = 1. \lab{normalisation_ABP1D} \eeq 
\end{subequations}

Conceptually, the ABP$^+$ problem \eqn{ABP1D} is simpler than its ABP counterpart: adding a spatial component makes the diffusion non-degenerate and leads to a smooth pdf without singular boundary terms. This makes the relationship between ABP$^+$ and ABP somewhat subtle. We now examine this relationship by carrying out an asymptotics analysis of ABP$^+$ for $\eps \to 0$. We start with the distinguished limit \(v=O(1)\). Our analysis extends that of \cite{Berlyand_etal2020} by capturing the details of the turning-point region. We consider a second distinguished limit, \(v=O(\eps^{1/2})\),  in
\S\ref{sec:weak}.

\begin{figure}[t]
\centering
\includegraphics[width=0.49\linewidth]{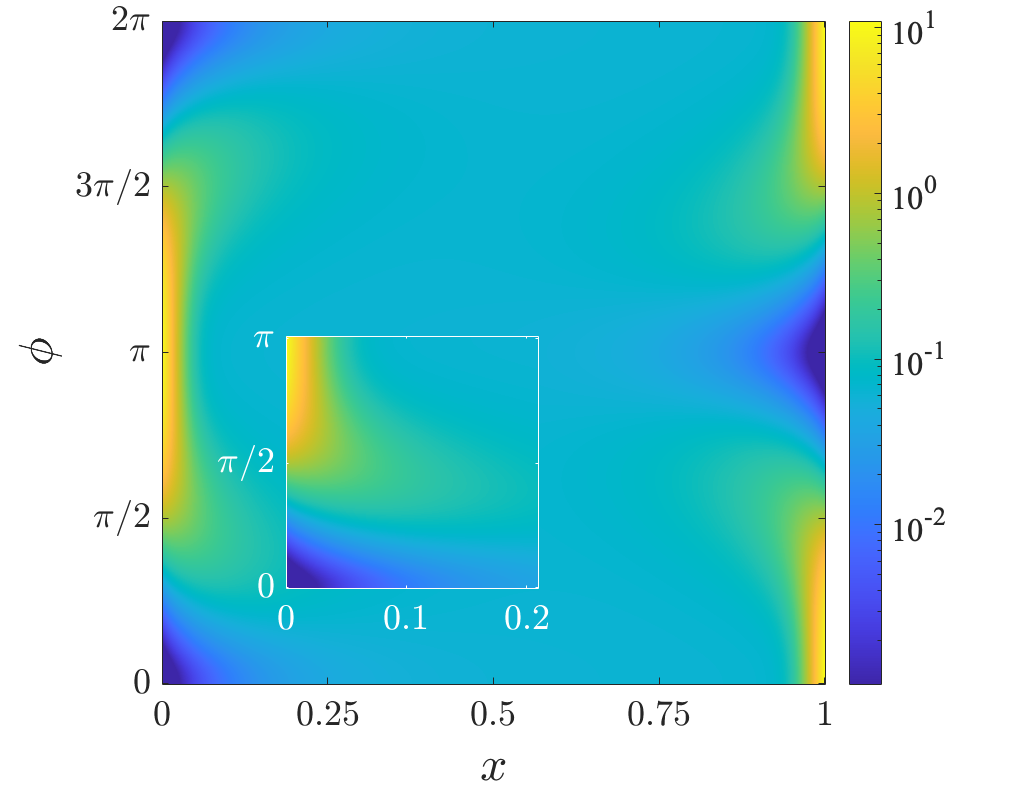}
  \hspace{0.01\linewidth}%
\includegraphics[width=0.49\linewidth]{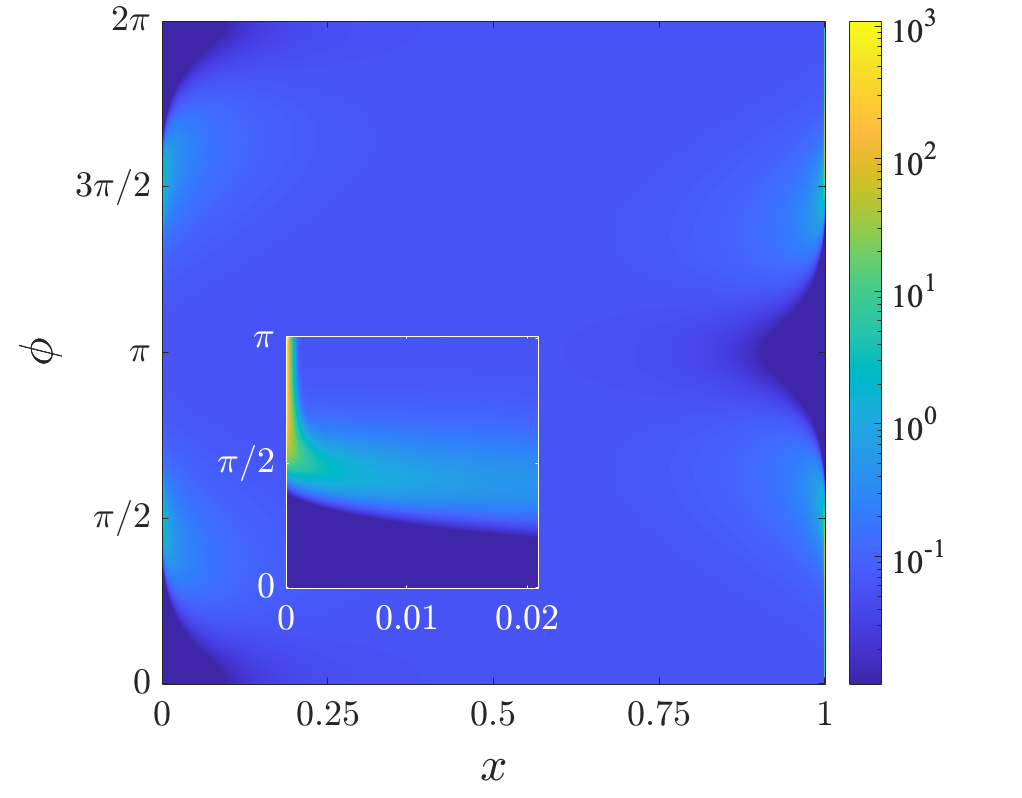}
  \caption{
Stationary pdf of ABP$^+$ obtained by solving
\eqn{ABP1D_tot} numerically as a function of position $x$
and orientation $\phi$ for $v=1$ and
$\eps=10^{-2}$ (left) and $\eps=10^{-4}$ (right).
The insets show close-ups of the turning-point region
near $(x,\phi)=(0,\pi/2)$. 
 }
  \label{fig:p_strong}
  \end{figure}

\subsection{Order-one swimming P\'eclet number, \(v=O(1)\)}

We begin by examining the stationary solution of \eqn{ABP1D} numerically.
Details of the numerical implementation are given in Appendix
\ref{sec:weakformulation}. Figure~\ref{fig:p_strong} compares the pdf $p(x,\phi)$ obtained for $v=1$ with \(\eps = 10^{-2}\) and $\eps=10^{-4}$. As \(\eps\) decreases,
pronounced boundary layers develop, together with localised structures
near the turning points.
We use matched asymptotics to capture these.
We focus on the left boundary \(x=0\), since the analysis at \(x=1\)
follows by symmetry.

\subsubsection{Bulk and boundary-layer solutions}\label{sec:MAE_boundary}

Away from the boundary, in the region $x=O(1)$ which we now regard as the bulk, the solution is approximated by the expansion  
\beq\lab{outerexp}
p = p^{(0)} +   O(\eps),  
\eeq
with the order of the error justified later on. 
Substituting \eqn{outerexp} into \eqn{ABP1D} we obtain  at $O(1)$ that  
\beq \lab{bulk}
 v \cos \phi \, p^{(0)}_x =  p^{(0)}_{\phi \phi}.  
\eeq

Near the boundary $x=0$ and away from the turning points $(0,\pi/2)$ and $(0,3\pi/2)$, there is a boundary layer of thickness $\eps$. We express the solution in the boundary layer using 
\beq\lab{boundary-layer-scaling}
X = x/\eps \quad \textrm{and} \quad P(X,\phi) = p(x,\phi),
\eeq
assuming $X=O(1)$.
  Eqs.\ \eqn{ABP1D} become
\begin{equation}\lab{inner}
\eps^{-1}v\cos\phi\,P_X
=
P_{\phi\phi}+\eps^{-1}P_{XX}
\quad \textrm{with} \ \
 P_X=v\cos\phi\,P
\quad\text{at} \ \  X=0.
\end{equation}
We expand the boundary-layer pdf as
\beq\lab{boundarylayer_exp}
P
=
\eps^{-1}P^{(0)}(X,\phi)
+
P^{(1)}(X,\phi)
+
O(\eps).
\eeq
The $O(\eps^{-1})$ leading-order term is motivated by the expectation that particles have an $O(1)$ probability to be in the boundary layer.
Substituting \eqn{boundarylayer_exp} into \eqn{inner} gives at orders $\eps^{-1}$ and $1$ 
\begin{subequations}
\begin{align}
v \cos \phi \, P^{(0)}_X &= P^{(0)}_{XX} \quad \textrm{with}  \ \ v \cos \phi \, P^{(0)} = P^{(0)}_{X} \ \ \textrm{at} \  X =0, \lab{a} \\
 v \cos \phi \, P^{(1)}_X &= P^{(0)}_{\phi \phi} + P^{(1)}_{XX},  \quad \textrm{with}  \ \  v \cos \phi \, P^{(1)} = P^{(1)}_{X} \ \ \textrm{at} \  X =0. \lab{b}
\end{align}
These equations are supplemented by the conditions 
\beq\lab{matching}
 P^{(0)} \to 0 \quad  \textrm{and} \quad
 P^{(1)} \to p^{(0)}(0,\phi)  
 \quad \textrm{as} \ \  X \to \infty
\eeq
\end{subequations}
ensuring matching with the bulk solution.
The leading-order solution is therefore given by 
\beq\lab{boundary layer form}
P^{(0)} = \begin{cases}
 0, & \textrm{for} \ \ \cos \phi > 0 \\
- v \cos \phi \AL(\phi) \e^{v \cos \phi \, X} & \textrm{for} \ \ \cos \phi < 0
\end{cases}.
\eeq
The factor $- v \cos \phi >0$ is introduced so that the so-far undetermined function $\AL(\phi)$ is interpreted as the (leading-order) probability for a particle to be in the boundary layer:
\beq\lab{mass}
\AL(\phi)  =\int_0^\infty P^{(0)}(X,\phi) \, \d X.
\eeq
This parallels the interpretation of $\AL(\phi)$ for the ABP model which replaces probability distributed over a narrow boundary layer by a singular wall contribution, cf.\ \eqn{total_PDF}.

 We  obtain an equation for $\AL(\phi)$   by integrating \eqn{b} from $X=0$ to $X \to \infty$ and using the  boundary condition together with the matching condition \eqn{matching}.  This gives
 \beq\lab{boundary1}
 \AL_{\phi \phi} = v \cos \phi \,  p^{(0)}(0,\phi)  
 \eeq
 for $\cos \phi < 0$. For $\cos \phi > 0$, we have $P^{(0)}=0$ and therefore $\AL=0$. 

The boundary-layer expansion is not uniform as
\(\cos\phi\to0\). Additional localised regions  arise near
the turning points \((0,\pi/2)\) and \((0,3\pi/2)\) to which we now turn.

 \begin{figure}[t]
\centering
\includegraphics[width=0.4\linewidth]
 {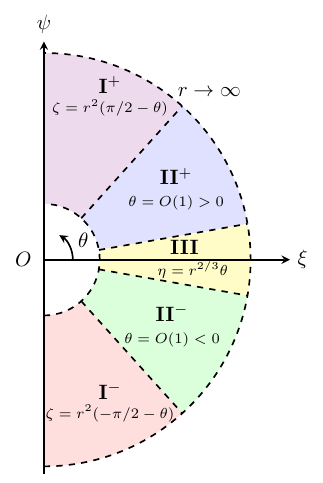}
\caption{
Schematic of the distinguished far-field regions of the
turning-point problem \eqn{cooleqn}. The inner semicircle
indicates that the near-field region \(r=O(1)\) is excluded. The
regions are matched in the order
\(\mathbf{I^-}\to\mathbf{II^-}\to\mathbf{III}\to\mathbf{II^+}\to\mathbf{I^+}\) (see Appendix \ref{app:turning_farfield}). 
}
\label{fig:turning_regions}
\end{figure}

\subsubsection{Turning-point regions}\label{sec:reentry}

The turning-point regions connect
the boundary-layer and bulk solutions and resolve the transition
from accumulation at the boundary for $\cos \phi < 0$ to re-entry into the bulk for $\cos \phi > 0$.
We consider the neighbourhood of \((x,\phi)=(0,\pi/2)\);
the other turning points are treated analogously.

Introducing
\begin{equation}\lab{turning_scaling}
\xi=\epsilon^{-3/4}x,
\quad
\psi=\epsilon^{-1/4}\left(\phi-{\pi}/{2}\right) \quad \textrm{and} \quad \P(\xi,\psi)=p(x,\phi),
\end{equation}
we rewrite the stationary problem as
\begin{equation}\lab{turning_rescaled_problem}
 -\epsilon^{-1/4}v\sin(\epsilon^{1/4}\psi)\P_\xi
 =\P_{\psi\psi}+\P_{\xi\xi} \quad \textrm{with} \ \ 
 \P_\xi=-\epsilon^{-1/4}v\sin(\epsilon^{1/4}\psi)\P
 \ \ \textrm{at} \ \ \xi=0.
 \end{equation}
Since 
\(\epsilon^{-1/4}\sin(\epsilon^{1/4}\psi)\)
has a Taylor expansion in powers of \(\epsilon^{1/2}\), we seek a solution
of the form
\begin{equation}\lab{expansion_sourcereg}
\P
=\epsilon^{-5/8}\P^{(0)}
+O(\epsilon^{-1/8}).
\end{equation}
The prefactor $\eps^{-5/8}$ is chosen to ensure matching with the bulk and boundary-layer solutions as we demonstrate below. 
Introducing \eqn{expansion_sourcereg} into \eqn{turning_rescaled_problem} gives the leading-order problem
\begin{equation}\lab{cooleqn}
-v\psi\,\P^{(0)}_\xi
=\P^{(0)}_{\psi\psi}+\P^{(0)}_{\xi\xi}
\quad \textrm{with} \ \
\P^{(0)}_\xi=-v\psi\,\P^{(0)}
\ \ \textrm{at} \ \ \xi=0.
\end{equation}
 Solving this problem with suitable matching with the bulk and boundary-layer solutions would provide a complete description of the capture--re-entry balance near the turning points. However, compared with the full stationary problem \eqn{ABP1D}, the sole simplification is the replacement of $\cos \phi$ by its linear approximation. An analytic solution therefore seems impractical.

For matching purposes, it is sufficient to determine the
far-field behaviour of $\P^{(0)}$. Using polar coordinates
$(r,\theta)$ in the $(\xi,\psi)$ plane, we obtain asymptotic
approximations as $r\to\infty$ in the five distinct 
regions shown in Figure~\ref{fig:turning_regions}.
The derivation is given in Appendix~\ref{app:turning_farfield}. We now describe the main features of the far-field solution. 

In region \(\mathbf{I^-}\),
 where \(\psi\to-\infty\),
\(\xi\to0^+\) and \(\xi\psi=O(1)\),
and in region \(\mathbf{II^-}\),  where
$\xi=O(\psi)$ as $\psi\to -\infty$, the solution is exponentially small. In the central region \(\mathbf{III}\), where
\(\xi\to\infty\) with \(\psi=O(\xi^{1/3})\),
the solution has the asymptotic form
\begin{equation}\lab{turning_Tricomi_ABP+}
\P^{(0)}(\xi,\psi)
=
\frac{\C}{4v}
\left(\frac{v}{9}\right)^{5/6}
\xi^{-5/6}\,
\F\left(\frac{\psi^3}{\xi}\right)
+O(\xi^{-13/6}),
\end{equation}
where \(\F\) is the Tricomi function obtained in
the turning-point asymptotics of ABP and given in \eqn{F}.
In region \(\mathbf{II^+}\), where
$\xi=O(\psi)$ as $\psi\to\infty$, 
the solution has the
asymptotic expansion
\begin{equation}
\P^{(0)}(\xi,\psi)
=
\frac{\C}{4v}\,\psi^{-5/2}
+O(\psi^{-9/2}).
\lab{turning_region_IIplus}
\end{equation}
The large-positive-argument limit of \(\F\) in
\eqn{turning_Tricomi_ABP+} recovers this behaviour in the
overlap between regions \(\mathbf{III}\) and
\(\mathbf{II^+}\).
When expressed in terms of $(x,\phi)$, \eqn{turning_Tricomi_ABP+} and \eqn{turning_region_IIplus} recover the singular behaviour of the ABP solution near the turning point given in \eqn{consequence} and \eqn{ABP0_turning_bulkboundary}.

In region
\(\mathbf{I^+}\), where \(\psi\to+\infty\),
 \(\xi\to0^+\) and \(\xi\psi=O(1)\),
 \eqn{regionI_exp} and \eqn{Q0} give
\begin{equation}\lab{regionIplus_higher}
\P^{(0)}(\xi,\psi)
=
v\,\C\,\psi^{3/2}\,\e^{-v\xi\psi}
+O(\psi^{-5/2}).
\end{equation}
%
This matches the leading-order solution in the  boundary layer.
Integrating \eqn{regionIplus_higher} over $\xi\in(0,\infty)$ gives
$\C\,\psi^{1/2}$, recovering
the square-root behaviour of $\AL$ in \eqn{assumption}
with $\C=\C_{\mathrm{I}^+}/v$.

The error terms in \eqn{turning_Tricomi_ABP+}
and 
\eqn{turning_region_IIplus}
match consistently with $O(\eps)$ errors in the bulk. 
In region \(\mathbf{I^+}\), 
the $O(\psi^{-5/2})$ term
 in \eqn{regionIplus_higher} can be matched to the higher-order boundary-layer term $P^{(1)}$ in \eqn{boundarylayer_exp}, leaving an 
$O(\eps)$
 error in the boundary. This matching follows from the asymptotic results \eqn{regionI_exp}, \eqn{Q1} and \eqn{C1plus_relation};
 the corresponding calculation of $P^{(1)}$ is not shown.

\begin{figure}[t]
 \centering
\includegraphics[width=0.49\linewidth]
  {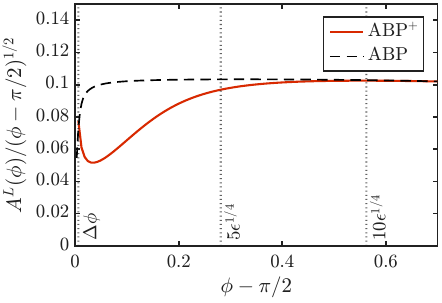}%
\hspace{0.01\linewidth}%
\includegraphics[width=0.49\linewidth]
  {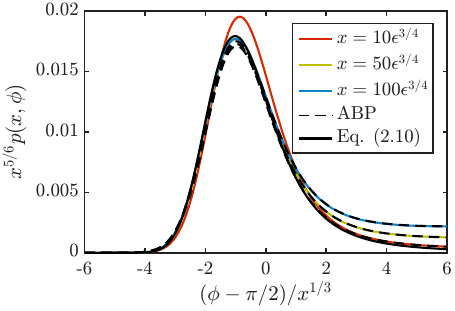}
\caption{
Comparison of the stationary solutions for ABP$^{+}$ and ABP in  the turning-point region near $(x,\phi)=(0,\pi/2)$, with $v=1$ and $\eps=10^{-5}$. The solutions are obtained numerically from \eqn{ABP1D_tot} and \eqn{MC}.
Left: boundary density rescaled by $(\phi-\pi/2)^{1/2}$ for ABP$^{+}$ (red solid curve) and ABP (black dashed curve).
Right: rescaled bulk density $x^{5/6}p(x,\phi)$ plotted against $(\phi-\pi/2)/x^{1/3}$  at selected values of $x=K\eps^{3/4}$ for ABP$^{+}$ (coloured solid curves) and ABP (black dashed curves), compared with a best-fit multiple of the similarity solution $\F$ in \eqn{F} (black solid curve).
}
\label{fig:constant_extraction}
\end{figure}

\subsubsection{ABP$^+$ vs ABP}\label{sec:recoveryABP}

We have shown that, in the regime $\eps \to 0, \, v=O(1)$, the stationary solution of the ABP$^+$ model matches exactly the behaviour \eqn{consequence} and \eqn{ABP0_turning_bulkboundary} of the ABP solution. The ABP model can therefore be thought of as the $\eps \to 0$ limit of the ABP$^+$ model. In the limiting process, the turning-point regions shrink to points and the probability distributed across the boundary layer concentrates on the boundary. At leading order, the details of the solution in the turning-point regions obtained from \eqn{cooleqn} do not affect the bulk and boundary layer solutions. Importantly, the probability for particles to be in the turning-point regions goes to $0$ as $\eps \to 0$, specifically as $\eps^{3/8}$ (since $\P \, \d x \d \phi = \eps^{3/8} \P^{(0)} \, \d \xi \d \psi$).
Restoring the dependence on the swimming P\'eclet number, this probability scales as
$O((\eps/v^2)^{3/8})$. It ceases to be small when
\(v=O(\eps^{1/2})\), signalling the transition to the weak-swimming regime described in Section~\ref{sec:weak} below.

We compare stationary solutions in the ABP$^+$ and ABP models for finite, small $\eps$ in
Figure~\ref{fig:constant_extraction}. For \(v=1\) and
\(\eps=10^{-5}\), a numerical solution of the  ABP\(^{+}\) problem
\eqn{ABP1D} is obtained and integrated over the boundary layer
\(0\leq x\leq10\eps\) to estimate the left-boundary density $\AL$. The 
ABP solution is obtained from the lattice Markov model \eqn{MC} in appendix \ref{app:MC}. 
Panel~(a) shows an excellent agreement between $\AL$ obtained for ABP and ABP\(^{+}\), both exhibiting the square-root behaviour \eqn{assumption} as $\phi \to \pi/2^+$ (provided that $\phi-\pi/2 \gg  \eps^{1/4}$ for ABP$^+$).
Panel~(b) shows the pdf $p(x,\phi)$ for ABP$^+$ for several values of $x$ in the range \(\eps^{3/4}\ll x\ll1\) where the bulk and turning-point regions overlap, together with the ABP  and  Tricomi similarity
solutions \eqn{ABP0_turning_bulkboundary}. The closest agreement with the similarity  solution is for
\(x=10\eps^{3/4}\simeq1.78\times10^{-3}\); at larger \(x\), the ABP and
ABP\(^{+}\) solutions remain essentially indistinguishable but both
depart from the similarity solution, as expected from the local nature of the similarity approximation.

 \begin{figure}[t]
    \centering
    \includegraphics[width=\textwidth]{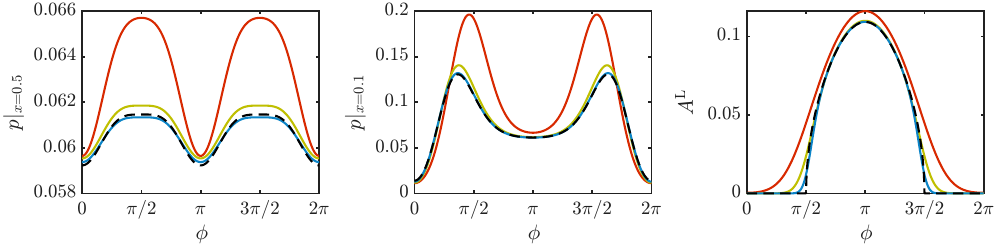}
    \caption{
Same as  Fig.~\ref{fig:constant_extraction} but comparing ABP$^+$ (solid curves) and ABP (black dashed curves)  over the full angular range $0\leq\phi\leq2\pi$: bulk pdf at $x=0.5$ (left) and $x=0.1$ (middle), and boundary density (right). The ABP$^+$ solutions correspond to $\eps=10^{-2}$ (red), $10^{-3}$ (yellow), and $10^{-4}$ (blue).
    }
    \label{fig:reduced_full_comparison}
\end{figure}

Figure~\ref{fig:reduced_full_comparison} compares numerical solutions
of  ABP\(^{+}\) and  \eqn{ABP1D} over the full range $0 \le \phi < 2 \pi$ for \(v=1\) and several
values of \(\eps\). The agreement between the ABP\(^{+}\) and ABP solutions
improves systematically as \(\eps\) decreases and is excellent for the
smallest value shown, confirming that ABP\(^{+}\)  
converges to ABP  as $\eps \to 0$. Our numerical estimate of the difference in $p$ between  
ABP\(^{+}\) and ABP (not shown) is consistent with the $O(\eps)$ error term in \eqn{outerexp}.

\subsection{Small swimming P\'eclet number, \(v=O(\eps^{1/2})\)} 
\label{sec:weak}

The convergence of ABP\(^{+}\) to ABP obtained for \(v=O(1)\) is not uniform
as the swimming P\'eclet number decreases: when
\(v=O(\eps^{1/2})\) the ABP reduction breaks down. 
We therefore
return to the full ABP\(^{+}\) problem and set
$v=\eps^{1/2}\bar v$.
A regular outer expansion in powers of \(\eps^{1/2}\)  shows that
the leading-order bulk pdf is independent of  \(x\) and
\(\phi\).
Unit normalisation therefore gives
$p^{(0)}={1}/{2\pi}$.
The boundary condition in \eqn{ABP1D} is restored in layers of thickness
\(O(\eps^{1/2})\). Near \(x=0\), introducing
\(\bar X=x/\eps^{1/2}\) and $p(x,\phi)=\bar P^{(0)}(\bar X,\phi) + O(\eps^{1/2})$, the leading boundary-layer problem is
\begin{equation}\lab{weakBLfinal}
\bar v\cos\phi\,\bar P^{(0)}_{\bar X}
=
\bar P^{(0)}_{\phi\phi}
+\bar P^{(0)}_{\bar X\bar X}
\quad \textrm{with} \  \
\bar P^{(0)}_{\bar X}
=
\bar v\cos\phi\,\bar P^{(0)} \ \
\textrm{at} \ \ \bar X=0, \ \
\end{equation}
with the matching condition $
\bar P^{(0)}\to 1/{2\pi}$  as $\bar X\to\infty$. Unlike in the $v=O(1)$ regime, the probability for particles to be in the boundary layer is negligible
and there is no separate turning-point scaling at leading order.
For \(\bar v\ll1\), \eqn{weakBLfinal} gives explicitly
\begin{equation}\lab{weak_bdy_explicit}
\bar P^{(0)}
\sim
\frac{1}{2\pi}
\left(1-\bar v\cos\phi\, \e^{-\bar X}\right).
\end{equation}

\begin{figure}
    \centering
    \begin{minipage}{.45\textwidth}
        \centering
        \includegraphics[width=\linewidth]
        {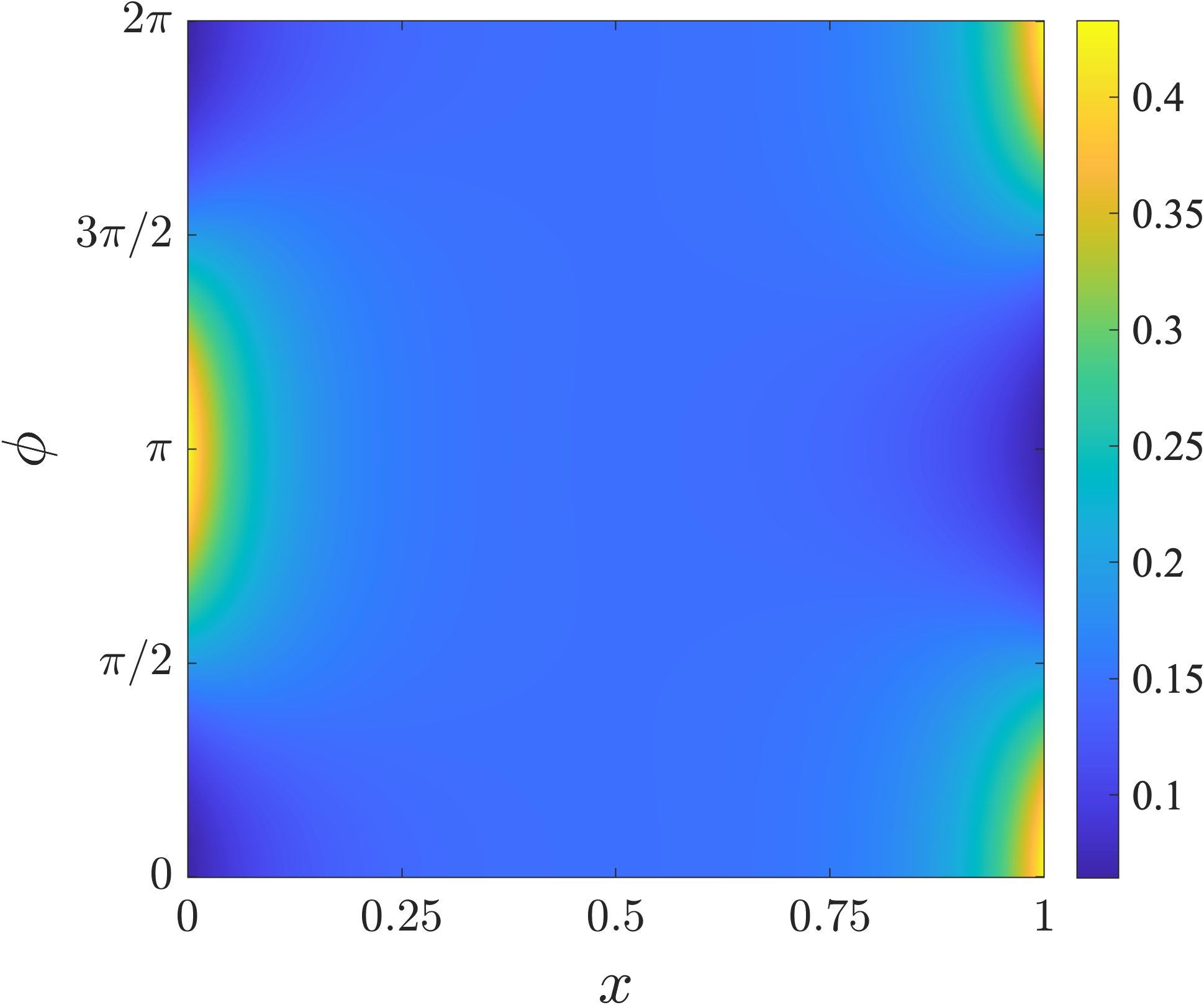}
    \end{minipage}%
   \hspace{0.1\linewidth}%
    \begin{minipage}{0.37\textwidth}
        \centering
        \includegraphics[width=\linewidth]
        {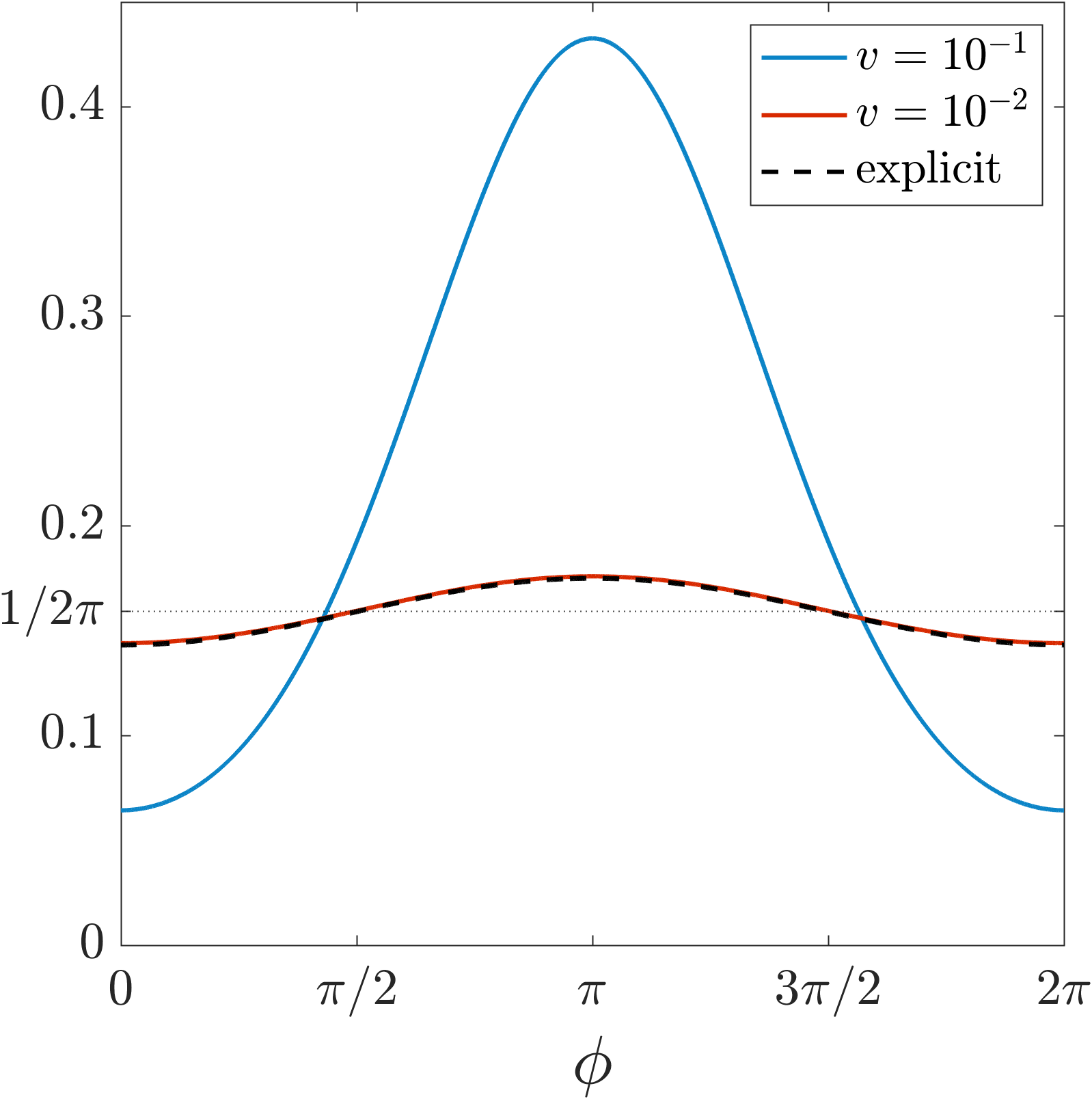}
    \end{minipage}
    \caption{
    Left: Stationary pdf $p(x,\phi)$ for
\(v=\eps^{1/2}=0.1\) and \(\eps=0.01\) obtained numerically by solving \eqn{ABP1D_tot}. 
Right: $p|_{x=0}(\phi)$  for different values
of \(v\) (thick solid lines), compared with the
 weak-swimming approximation \eqn{weak_bdy_explicit}
at \(\bar X=0\) (dashed line).
    }
    \label{fig:p_weak}
\end{figure}

Figure~\ref{fig:p_weak} confirms these conclusions.
Panel (a) shows the numerical solution of \eqn{ABP1D} for
\(\eps=0.01\) and \(v=\eps^{1/2}=0.1\). The solution is nearly uniform in the bulk, with orientation dependence confined to boundary layers of thickness \(O(\eps^{1/2})\), as predicted. Panel (b) shows the solution on the left boundary $x=0$ for \(v=\eps^{1/2}=0.1\) and \(v=\eps=0.01\). In both cases the concentration is larger for orientations directed towards the wall, but as the swimming speed is reduced the boundary modulation weakens and the numerical solution approaches the explicit approximation \eqn{weak_bdy_explicit}.

\section{Conclusions}\label{sec:conclusions}

This paper provides a detailed description of the stationary distribution of ABPs in a channel, first without spatial diffusion (ABP model), then with a weak spatial diffusivity $\eps$ (ABP$^+$ model). A key result is the similarly solution \eqn{self-similar} for the bulk density near the turning points, with $\F$ expressed in terms of a Tricomi confluent hypergeometric function \eqn{F}. This shows, in particular, that for $\eps = 0$ the bulk density is singular on the boundary, with $p|_{x=0} \propto (\phi - \pi/2)^{-5/2}$ as $\phi \to \pi/2^+$. The boundary density scales as $\AL \propto (\phi - \pi/2)^{1/2}$ corresponding to a singular flux $\AL_\phi \propto (\phi - \pi/2)^{-1/2}$. A conclusion is that the exchanges between bulk and boundary near the turning points cannot simply be interpreted as a straightforward balance between the incoming flux $v \cos \phi \, p_x$ and the outgoing flux $\AL_{\phi}$ since neither is finite. 

A finite diffusivity $\eps > 0$ regularises the problem: the bulk-boundary exchanges are distributed across a turning-point region characterised by $x=O(\eps^{3/4})$ and $|\phi-\pi/2| = O(\eps^{1/4})$ in which the bulk density scales as $p=O(\eps^{-5/8})$. It is worth noting that other regularisations are also relevant: in particular, spatial discretisation as in the Markov lattice model used for numerical computation provides an alternative regularisation, with the lattice spacings $\Delta x$ and  $\Delta \phi \to 0$ playing the role of regularisation parameters.

 A natural next step is to study the time-dependent ABP model. Fu, Perthame and Tang~\cite{FuPerthameTang2021} established weak convergence of solutions of the time-dependent ABP system \eqn{ABP_PDE} to a steady state, but the local behaviour near the turning points remains unresolved, including whether the singular structure develops at finite times or only in the long-time limit. An important question is therefore how the  turning-point structure identified here develops in time and how the time evolution is regularised by small spatial diffusivity.

The analysis carried out in this paper can be extended to general smooth boundaries. Experiments show that curvature can strongly redistribute bacterial accumulation and promote trapping in highly curved regions~\cite{Mok_etal2019,PerezEstay_etal2024}. Addressing such geometries requires restoring the spatial dependence tangential to the boundary that is absent from the spatially one-dimensional reduction considered here. In this more general setting, curvature enters both the bulk and boundary equations, as in the analysis of Berlyand et al.~\cite{Berlyand_etal2020} for small spatial diffusivity. The leading turning-point structure is expected to remain locally that obtained here for a flat boundary, while curvature may redistribute the boundary-supported probability and produce spatially localised accumulation.

Finally, the point-particle description used here neglects the orientation-dependent geometric constraints associated with finite particle size and shape. Chen and Thiffeault~\cite{ChenThiffeault2021} showed that these can strongly influence the stationary distribution of an active Brownian particle in a channel in the limit of weak rotational diffusion, where the swimming persistence length is large compared with the channel width and spatial diffusion remains at leading order. It remains to understand how finite size and shape affect the contrasting regime of small spatial diffusivity  considered here, particularly when the persistence length is comparable to the channel width and the limiting distribution develops a singular boundary-supported component.

\section*{Acknowledgements}

A. Tzella   is grateful to  J. Mike F. Gunn and  Jean-Luc Thiffeault   for  helpful discussions related to this work.
She was partially supported by an Emmy Noether Fellowship from the London Mathematical Society.

\appendix
\numberwithin{equation}{section}
\renewcommand{\theequation}{\Alph{section}\arabic{equation}}

\section{Lattice Markov chain discretisation of the ABP model} \label{app:MC}

The Markov chain model is obtained by discretising the domain
$(x,\phi)\in[0,1]\times[0,2\pi]$ into $M$ bulk cells,
with centres $x_i$ and widths $\Delta x_i$,
$i=1,\ldots,M$.
The angular domain is discretised into $N$ sites
$\phi_j=j\Delta\phi$, where $\Delta\phi=2\pi/N$
and $N$ is a multiple of $4$.
Boundary states are retained only for orientations directed towards
 the boundary: $\cos\phi_j<0$ at $x=0$ and $\cos\phi_j>0$ at $x=1$.

Particles in bulk cell $(i,j)$ hop to $(i,j+1)$ and $(i,j-1)$
with rate $(\Delta\phi)^{-2}$ and, according to the sign of
$\cos\phi_j$, to the next bulk cell or the corresponding boundary
state with rate $v|\cos\phi_j|/\Delta x_i$.
Particles on boundary states undergo angular hopping at the same
rate $(\Delta\phi)^{-2}$.
An angular hop that reaches a turning point  $\cos\phi_j=0$
(corresponding to $j=N/4$ or $3N/4$) transfers the particle to
the nearest bulk cell, with $i=1$ at the left wall and $i=M-1$
at the right wall.

We denote by $p_{i,j}(t)$, with $i=1,\ldots,M$ and
$j=0,\ldots,N-1$, the probability that a particle occupies a bulk cell $(i,j)$,
and by $\AL_j(t)$ and $\AR_j(t)$ the probabilities 
that a particles occupies a left
or right boundary state, respectively.
The forward Kolmogorov equations are
\begin{subequations}
\lab{MC}
\begin{align}
\dot p_{i,j}
&=\frac{p_{i,j+1}+p_{i,j-1}-2p_{i,j}}{(\Delta\phi)^2}
+v|\cos\phi_j|
 \left(\frac{p_{i-\sigma_j,j}}{\Delta x_{i-\sigma_j}}
       -\frac{p_{i,j}}{\Delta x_i}\right)
&&\text{for }\cos\phi_j\ne0, \lab{MC1}\\
\dot{A}^{\textsc{l}}_j
&=\frac{\AL_{j+1}+\AL_{j-1}-2\AL_j}{(\Delta\phi)^2}
  -v\cos\phi_j\frac{p_{1,j}}{\Delta x_1}
&&\text{for }\cos\phi_j<0, \lab{MC3}\\
\dot{A}^{\textsc{r}}_j
&=\frac{\AR_{j+1}+\AR_{j-1}-2\AR_j}{(\Delta\phi)^2}
  +v\cos\phi_j\frac{p_{M,j}}{\Delta x_M}
&&\text{for }\cos\phi_j>0. \lab{MC4}
\end{align}
where $\sigma_j=\operatorname{sgn}(\cos\phi_j)$ 
and the angular index $j$ is interpreted modulo $N$.
Eq.\ \eqn{MC1} 
 uses $p_{0,j}=0$   when $\sigma_j=1$
and  $p_{M,j}=0$ when $\sigma_j=-1$.
Similarly, equations  \eqn{MC3}--\eqn{MC4} use  $\AL_j=\AR_j=0$ at 
 the turning points $j=N/4$ and
$j=3N/4$. 
These zero values
do not represent additional Markov-chain states.
To complete \eqn{MC}, the bulk probabilities for $\cos \phi_j=0$ satisfy
\begin{equation}
\dot p_{i,j}
=\frac{p_{i,j+1}+p_{i,j-1}-2p_{i,j}}{(\Delta\phi)^2}
 +\delta_{i,1}\frac{\AL_{j\pm1}}{(\Delta\phi)^2}
 +\delta_{i,M-1}\frac{\AR_{j\mp1}}{(\Delta\phi)^2},
\qquad \text{for }\cos\phi_j=0.
\lab{MC5}
\end{equation}
\end{subequations}
Here $\delta_{i,k}$ is the Kronecker delta, and the upper and lower
signs correspond to $j=N/4$ and $j=3N/4$, respectively.
The last two terms in \eqn{MC5} account for re-entry.
Each gain equals the loss from the adjacent boundary state hopping
towards the turning orientation; these losses are already included
in the angular second differences in \eqn{MC3}--\eqn{MC4}.

We note that the marginal probability obtained by
summing over the spatial indices satisfy
\begin{equation}
\dot m_j=\frac{m_{j+1}+m_{j-1}-2m_j}{(\Delta\phi)^2},
\qquad
m_j=\sum_{i=1}^{M-1}p_{i,j}+\AL_j+\AR_j.
\lab{MC_angular_marginal}
\end{equation}
Summing over $j$ shows that the complete system \eqn{MC}
conserves total probability:
\begin{equation}
\sum_{j=0}^{N-1}\sum_{i=1}^{M-1}p_{i,j}
+\sum_{j=0}^{N-1}\AL_j
+\sum_{j=0}^{N-1}\AR_j= 1,
\lab{MC_normalisation}
\end{equation}
where  we have extended  
$\AL_j$ and $\AR_j$ by zero where their states
do not exist. 
System \eqn{MC} is a standard forward Kolmogorov equation for a continuous-time Markov chain.
A system equivalent to \eqn{MC} is obtained by \cite{FuPerthameTang2021} as a discretisation of \eqn{ABP_PDE} on a uniform lattice so that $\Delta x_i=\Delta x/M$.

We compute the stationary distribution directly by solving the linear system obtained by setting the left-hand side of \eqn{MC} to $0$ and imposing the
normalisation \eqn{MC_normalisation}.
We choose variable lattice  spacing 
$\Delta x_i$ to resolve the turning-point structure.
We take $M$ even and $N=8M$. The cell centres are
$x_i=2((i-1)^3+i^3)/M^3$ for $i=1,\ldots,M/2$, with
$x_{M+1-i}=1-x_i$. The corresponding widths are
$\Delta x_i=4(i^3-(i-1)^3)/M^3$, with
$\Delta x_{M+1-i}=\Delta x_i$.

We now derive a weak formulation of the stationary system \eqn{MC} before passing
formally to the continuum limit. Let $q(x,\phi)$ be a sufficiently
regular test function, $2\pi$-periodic in $\phi$. At the bulk and boundary
states, set $q_{i,j}=q(x_i,\phi_j)$ for $1\leq i\leq M$, $q_{0,j}=q(0,\phi_j)$ and
$q_{M+1,j}=q(1,\phi_j)$.
At stationarity, multiply every bulk equation, including \eqn{MC5},
by $q_{i,j}$, the left-boundary equations \eqn{MC3} by $q_{0,j}$, and
the right-boundary equations \eqn{MC4} by $q_{M+1,j}$, and sum over all
states. This yields the exact discrete
weak formulation
\begin{align}
0={}&\sum_{i=1}^{M}\sum_{j=0}^{N-1}p_{i,j}
\left(
\frac{q_{i,j+1}-2q_{i,j}+q_{i,j-1}}{(\Delta\phi)^2}
+\frac{v|\cos\phi_j|}{\Delta x_i}
(q_{i+\sigma_j,j}-q_{i,j})
\right) \nonumber \\
&+\frac{1}{(\Delta\phi)^2}\sum_{j=0}^{N-1}
\left(\AL_j(q_{0,j+1}-2q_{0,j}+q_{0,j-1})
+\AR_j(q_{M+1,j+1}-2q_{M+1,j}+q_{M+1,j-1})\right)  \nonumber\\
&+\frac{1}{(\Delta\phi)^2}\left(
\AL_{N/4+1}(q_{1,N/4}-q_{0,N/4})
+\AL_{3N/4-1}(q_{1,3N/4}-q_{0,3N/4})\right.  \nonumber \\
&\hspace{15mm}\left.
+\AR_{N/4-1}(q_{M,N/4}-q_{M+1,N/4})
+\AR_{3N/4+1}(q_{M,3N/4}-q_{M+1,3N/4})\right).
\lab{MC_discrete_weak}
\end{align}
The final bracket  accounts for re-entry. 
Substituting $p_{i,j}\simeq p(x_i,\phi_j)\Delta x_i\Delta\phi$,
and
$\AL_j\simeq\AL(\phi_j)\Delta\phi$,
$\AR_j\simeq\AR(\phi_j)\Delta\phi$  into \eqn{MC_discrete_weak}, we  pass formally to the continuum limit $\max_i\Delta x_i\to0$ and
$\Delta\phi\to0$.
The bulk and boundary sums give the corresponding integrals in \eqn{eq:FV_continuum_weak}.
The
re-entry terms vanish since $\AL_{N/4+1}$ and its counterparts tend to zero 
by continuity
of $\AL$ and using $\AL(\pi/2)=0$.

\section{Finite-element discretisation of the ABP$^+$ model}\label{sec:weakformulation}

We derive a weak formulation of the stationary ABP$^+$ problem
\eqn{ABP1D}.
Let $q\in H^1\big((0,1)\times(0,2\pi)\big)$ be a test function that is $2\pi$-periodic in $\phi$. 
Multiplying
\eqn{ABP1D} by \(q\) and integrating by parts
gives
\begin{equation}\lab{weak1}
\int_0^1\int_0^{2\pi}
\left(
p_\phi q_\phi
+\eps p_x q_x
-v\cos\phi\,p\,q_x
\right)
\,\d\phi\,\d x
=0.
\end{equation}
The spatial boundary terms cancel by the no-flux condition in
\eqn{ABP1D}, while the angular boundary terms cancel by
\(2\pi\)-periodicity.

We discretise \eqn{weak1} using a conforming piecewise-linear
finite-element method on a triangular mesh of
\([0,1]\times[0,2\pi]\), generated using MATLAB's PDE Toolbox. The mesh
is quasi-uniform in the bulk and   geometrically graded towards
\(x=0\) and \(x=1\) in order to resolve the \(O(\eps)\) boundary
layers. The grading is applied within a distance \(10\eps\) of each
boundary, with the element sizes decreasing exponentially towards the
boundary. The nodes at \(\phi=0\) mirror those at \(\phi=2\pi\), to allow for periodicity to be imposed exactly.
The normalisation \eqn{normalisation_ABP1D} is imposed using a
Lagrange multiplier. Element integrals are evaluated
using a third-order Gaussian quadrature rule. The resulting sparse
  linear system is solved using MATLAB's sparse direct solver. The
finite-element solution is subsequently interpolated onto a regular
\((x,\phi)\)-grid for visualisation and post-processing.

\section{Far-field analysis of the turning-point region}
\label{app:turning_farfield}

In this appendix we derive the far-field asymptotic structure of
the solution \(\P^{(0)}\) of \eqn{cooleqn}. 
We introduce the polar coordinates
\begin{equation}\lab{polar_trans}
\xi=r\cos\theta,
\qquad
\psi=r\sin\theta
\quad \textrm{for} \ \
r>0 \ \ \textrm{and} \ \
-\pi/2<\theta<\pi/2.
\end{equation}
The far-field limit corresponds to \(r\to\infty\).  The polar representation,  
$
\Q(r,\theta)=\P^{(0)}(\xi,\psi),
$
satisfies
\begin{subequations}\lab{cooleqn_polar}
\begin{align}
\Q_{rr}
+\frac{1}{r}\Q_r
+\frac{1}{r^2}\Q_{\theta\theta}
&=
-vr\sin\theta\cos\theta\,\Q_r
+v\sin^2\theta\,\Q_\theta,
\lab{polar_eqn}
\\
\Q_\theta
&=
vr^2\Q
\quad \textrm{at} \ \
 \theta=\pm\frac{\pi}{2}.
\lab{polar_bc}
\end{align}
\end{subequations}
Examination of the
dominant balances in \eqn{polar_eqn} identifies five distinguished
regions:
\begin{subequations}\label{regions}
\begin{align}
\textbf{Region }\mathbf{I^{\pm}}:&\quad
\zeta=r^2(\pm\pi/2-\theta)=O(1)
\quad \textrm{for}
\ \ \zeta\gtrless0, \lab{regionI}
\\
\textbf{Region }\mathbf{II^{\pm}}:&\quad
\theta=O(1)
\quad \textrm{for} \ \
\theta\gtrless0,
\\
\textbf{Region }\mathbf{III}:&\quad
\eta=r^{2/3}\theta=O(1), \lab{regionIII}
\end{align}
\end{subequations}
as \(r\to\infty\). 
The regions are shown schematically in
Figure~\ref{fig:turning_regions}.  We analyse the regions in the order
\(\mathbf{I^-}\), \(\mathbf{II^-}\), \(\mathbf{III}\),
\(\mathbf{II^+}\) and \(\mathbf{I^+}\), matching the solution in sequence.
Eqs.\ \eqn{regionI}--\eqn{regionIII} identify the variables $\zeta$, $\theta$ and $\eta$ used in conjunction with $r$ in each region. In what follows we abuse notation by using the same symbol $\Q$ for  $\P^{(0)}$ whatever the independent variables.

The near-boundary regions  \(\mathbf{I^-}\) and  \(\mathbf{I^+}\) are distinguished by signs:
 the upper (lower) sign in \eqn{regionI} corresponds to regions \(\mathbf{I^+}\) (
 \(\mathbf{I^-}\)) where \(\zeta>0\) (\(\zeta<0\)). 
 In both regions,
 \(\sin\theta\cos\theta=\zeta/r^{2}+O(r^{-6})\) and
 \(\sin^{2}\theta=1+O(r^{-4})\), so 
 the two regions are described by the
 same equation, 
 \begin{equation}\lab{cooleqn_polar_Region1}
 \Q_{\zeta\zeta}+v\Q_{\zeta}
 +r^{-2}\Q_{rr}
 +r^{-3}\left((1+v\zeta)\Q_{r}+4\zeta \Q_{r\zeta}\right)
 +r^{-4}\left(
 4\zeta^{2}\Q_{\zeta\zeta}
 +\left(4\zeta+v\zeta^{2}\right)\Q_{\zeta}
 \right)
 =O(r^{-7})
 \end{equation}
 obtained from \eqn{polar_eqn}.
The associated boundary condition obtained from \eqn{polar_bc} is
 \begin{equation}\lab{Region1_bc}
 \Q_{\zeta}=-v\Q
 \qquad\text{at}\quad\zeta=0 .
 \end{equation}
The solution in region \(\mathbf{I^-}\) needs matching with the boundary-layer solution which vanishes to all algebraic orders. It  therefore also vanishes to all orders: 
$
\Q(r,\zeta)
=
o(r^{-N})
\quad
\text{for every }N>0$.

In region \(\mathbf{II^-}\), 
we seek a solution that is also beyond all orders, specifically in the  form
\begin{equation}
\Q(r,\theta)
=
\C_{\mathrm{II}^-} \,r^a
\exp\!\left(
vr^2F_0(\theta)+F_1(\theta)+O(r^{-2})
\right), \lab{solution_regionII}
\end{equation}
where $\C_{\mathrm{II}^-}$ and $a$ are constant and we
assume that $F_0(\theta) < 0$ for $-\pi/2 < \theta < 0$.
Substitution into \eqn{polar_eqn} shows that the dominant balance selects
the exponential scaling \(r^2\) and gives
\begin{align}
(F_0')^2
-\sin^2\theta\,F_0'
+4F_0^2
+2\sin\theta\cos\theta\,F_0
&=0,
\lab{F0_equation} \\
\left(2F_0^{\prime}-\sin^2\theta\right)F_1^{\prime}
+
F_0^{\prime\prime}
+
4(a+1)F_0
+
a\sin\theta\cos\theta
&=
0,
\lab{F1_equation}
\end{align}
at the first two orders.
The function $F_0$ is odd   and satisfies $F_0'(0)=0$.  It has the local expansion
\begin{equation}
F_0(\theta)
=
{\theta^3}/{9}
-{\theta^5}/{27}
+O(\theta^7)
\quad \textrm{as} \ \
\theta\to0^- .
\lab{smallthetaF0}
\end{equation}
Using this and \eqn{F0_equation} it can be shown that  \(F_0<0\) for $-\pi/2 < \theta < 0$ as assumed.
For subsequent matching with region \(\mathbf{III}\), the local behaviour of \(F_1\) is useful. Using
\eqn{F1_equation} and \eqn{smallthetaF0} we obtain
\begin{equation}
F_1(\theta)
=
(3a+2)\log(-\theta)
-\frac12\left(a+\frac{10}{9}\right)\theta^2
+O(\theta^4)
\quad \textrm{as} \ \
\theta\to0^-,
\lab{smallthetaF1}
\end{equation}
where an additive constant has been absorbed into
\(\C_{\mathrm{II}^-}\).

In region \(\mathbf{III}\), introducing $\Q = \Q(r,\eta)$
into \eqn{polar_eqn} and multiplying by
\(r^{2/3}\) gives
\begin{equation}
\Q_{\eta\eta}
-\frac{v}{3}\eta^2\Q_\eta
+v\eta\,rQ_r
=
O(r^{-4/3}\Q),
\lab{eta_equation}
\end{equation}
at leading order. Here we make the assumptions 
\(r\Q_r=O(\Q)\), \(r^2Q_{rr}=O(\Q)\) and
\(r\Q_{r\eta}=O(\Q_\eta)\) which are verified a posteriori.
In the overlap with region \(\mathbf{II^-}\),
\eqn{smallthetaF0} and \eqn{smallthetaF1}, together with \eqn{solution_regionII}, give
\(
\Q\sim
\C_{\mathrm{II}^-}\,r^a(-\theta)^{3a+2}
\, \e^{vr^2\theta^3/9}
=
\C_{\mathrm{II}^-}\,r^{-a-4/3}(-\eta)^{3a+2}
\, \e^{v\eta^3/9}.
\)
Writing \(b=-a-4/3\), we therefore seek an expansion of the form
\begin{equation}\lab{RegionIII_exp}
\Q(r,\eta)
=
r^b\Q_0(\eta)
+O(r^{b-4/3}).
\end{equation}
Substitution into \eqn{eta_equation} gives
\begin{equation}\lab{Kummer_app}
\Q_{0,\eta\eta}
-\frac{v}{3}\eta^2\Q_{0,\eta}
+bv\eta \Q_0
=
0.
\end{equation}
Eq.\ \eqn{Kummer_app} is identical to the Kummer equation obtained earlier in
\eqn{Kummer}. 
For \(\eta<0\), the algebraic solution would generate an algebraic
contribution in region \(\mathbf{II^-}\), which has already been excluded.
We therefore retain the exponentially recessive solution 
\begin{equation}\lab{Q0-}
\Q_0(\eta)
=
\C_{\rm III}^- \,
\e^{v\eta^3/9} \,
\U\!\left(
 b+\frac23,\frac23,-\frac{v\eta^3}{9}
\right)
\quad \textrm{for} \ \ \eta<0.
\end{equation}
Using similar arguments as in \S\ref{subsec:reentry}, we conclude that for \(\eta>0\), excluding exponential growth as \(\eta\to+\infty\) gives
\begin{equation}\lab{Q0+}
\Q_0(\eta)
=
\C_{\rm III}^+
\U\!\left(
-b,\frac23,\frac{v\eta^3}{9}
\right)
\quad \textrm{for} \  \ \eta>0,
\end{equation}
with 
$b=-5/6$ and 
$\C_{\rm III}^-=6\,\C_{\rm III}^+$.

For \(\eta\to-\infty\), using large-argument
asymptotics \(\U(a,b,z)\sim z^{-a}\)
\cite[Eq.~(13.7.3)]{DLMF} gives 
\begin{equation}\lab{largenegeta}
\Q_0(\eta)
\sim
6\,\C_{\rm III}^+
\left(\frac{v}{9}\right)^{1/6}
(-\eta)^{1/2} \,
\e^{v\eta^3/9}
\quad \textrm{as} \ \
\eta\to-\infty.
\end{equation}
Matching \eqn{largenegeta} with the
region-\(\mathbf{II^-}\) solution \eqn{solution_regionII}
determines 
\begin{equation}
a=-1/2 \quad \textrm{and} \quad 
\C_{\mathrm{II}^-}
=
6 \left(v/9\right)^{1/6} \,\C_{\rm III}^+
.
\end{equation}
For \(\eta\to+\infty\), the large-argument asymptotics of
$\U$ \cite[Eq.~(13.7.3)]{DLMF} gives 
\begin{equation}\lab{largeposeta}
\Q_0(\eta)
=
\C_{\rm III}^+
\left(\frac{9}{v}\right)^{5/6}
\eta^{-5/2}
\left(1-\frac{35}{4v}\eta^{-3}+O(\eta^{-6})\right)
\quad \textrm{as} \ \ 
\eta\to+\infty.
\end{equation}

In region \(\mathbf{II^+}\) where \(\theta=O(1)>0\), the requirement of matching with \eqn{largeposeta} 
suggests
the expansion
\begin{equation}\lab{RegionIIp_exp}
\Q(r,\theta)
=
r^{-5/2}\Q_0(\theta)
+r^{-9/2}\Q_1(\theta)+O(r^{-13/2}).
\end{equation}
Substitution into \eqn{polar_eqn} gives, at leading order,
\(
\sin^2\theta\,\Q_{0,\theta}
+
\frac52\sin\theta\cos\theta\,\Q_0
=
0.
\)
Hence
\(
\Q_0(\theta)
=
\C\sin^{-5/2}\theta.
\)
Matching with \eqn{largeposeta} determines the amplitude, giving
\begin{equation}\lab{Q0_regionIIp}
\Q_0(\theta)
=
\C_{\rm III}^+
\left(9/v\right)^{5/6}
\sin^{-5/2}\theta.
\end{equation}
At  \(O(r^{-9/2})\), \eqn{polar_eqn} gives
\(
\sin^2\theta\,\Q_{1,\theta}
+\frac92\sin\theta\cos\theta\,\Q_1
=
\frac{1}{v}\left(\Q_{0,\theta\theta}+\frac{25}{4}\Q_0\right).
\)
Matching with the second term in \eqn{largeposeta} yields
\begin{equation}\lab{Q1_regionIIp}
\Q_1(\theta)
=
-\frac{35}{4v}\,
\C_{\rm III}^+
\left({9}/{v}\right)^{5/6}
\cos\theta\,\sin^{-11/2}\theta.
\end{equation}

Finally in region \(\mathbf{I^+}\) governed, like region \(\mathbf{I^-}\), by \eqn{cooleqn_polar_Region1}, 
we seek an 
expansion of the form
\begin{equation}\lab{regionI_exp}
\Q(r,\zeta)
=
r^{3/2}\Q_0(\zeta)
+
r^{-5/2}\Q_1(\zeta)
+
O(r^{-13/2}).
\end{equation}
Substitution into \eqn{cooleqn_polar_Region1} gives, at leading order,
\begin{equation}
\Q_{0,\zeta\zeta}+v\Q_{0,\zeta}=0
\quad
\text{with}
\quad
\Q_{0,\zeta}=-v \Q_0
\quad\text{at}\ \ \zeta=0,
\end{equation}
requiring $\lim_{\zeta\to\infty}\Q_0(\zeta)=0$.
Hence,
\begin{equation}\lab{Q0}
\Q_0(\zeta)
=
\C_{\mathrm{I}^+}\, \e^{-v\zeta},
\end{equation}
where $\C_{\mathrm{I}^+}$ is a constant. At \(O(r^{-5/2})\), we then obtain
\begin{equation}
\Q_{1,\zeta\zeta}+v\Q_{1,\zeta}
=
-\C_{\mathrm{I}^+}
\left(
3v^2\zeta^2-\frac{17}{2}v\zeta+\frac94
\right)\, \e^{-v\zeta} \quad \textrm{with}
\ \
\Q_{1,\zeta}=-v\Q_1 \quad \textrm{at} \ \ \zeta = 0.
\end{equation}
The solution is 
\begin{equation}\lab{Q1}
\Q_1(\zeta)
=
\D \, \e^{-v\zeta}
+
\frac{\C_{\mathrm{I}^+}}{4v^2}
\left(
1-
\left(1+v\zeta+5v^2\zeta^2-4v^3\zeta^3\right) \, \e^{-v\zeta}
\right),
\end{equation}
where \(\D=\Q_1(0)\) remains undetermined at this order.
Matching  with region
\(\mathbf{II^+}\) imposes that $
\lim_{\zeta\to\infty}\Q_1(\zeta)
=
\C_{\rm III}^+
\left({9}/{v}\right)^{5/6}$, leading to 
\begin{equation}\lab{C1plus_relation}
\C_{\mathrm{I}^+}
=
4v^2\left({9}/{v}\right)^{5/6}\C_{\rm III}^+.
\end{equation}
This completes the leading
far-field matching across the turning-point region.

\bibliographystyle{unsrtnat}
\bibliography{active}

\end{document}